\documentclass[prd,superscriptaddress,twocolumn,showpacs,amsmath,%
preprintnumbers,showkeys]{revtex4}
\usepackage{bm}
\usepackage{graphicx}
\begin{document}                                           
\title{Pion transverse charge density from timelike form factor data}
\author{G.~A.~Miller}
\affiliation{University of Washington, Seattle, WA 98195--1560, USA}
\author{M.~Strikman}
\affiliation{Department of Physics, Pennsylvania State University,
University Park, PA 16802, USA}
\author{C.~Weiss}
\affiliation{Theory Center, Jefferson Lab, Newport News, VA 23606, USA}
\date{November 2, 2010}
\begin{abstract}
The transverse charge density in the pion can be represented as a 
dispersion integral of the imaginary part of the pion form factor in 
the timelike region. This formulation incorporates information from 
$e^+e^-$ annihilation experiments and allows one to reconstruct the
transverse density much more accurately than from the spacelike pion 
form factor data alone. We calculate the transverse density using an 
empirical parametrization of the timelike pion form factor and estimate
that it is determined to an accuracy of $\sim 10\%$ at a distance 
$b \sim 0.1 \, \textrm{fm}$, and significantly better at larger distances.
The density is found to be close to that obtained from a zero--width 
$\rho$ meson pole over a wide range and shows a pronounced rise at 
small distances. The resulting two--dimensional image of the fast--moving 
pion can be interpreted in terms of its partonic structure in QCD.
We argue that the singular behavior of the charge density at the center 
requires a substantial presence of pointlike configurations in the pion's 
partonic wave function, which can be probed in other high--momentum 
transfer processes.
\end{abstract}
\keywords{Pion form factor, dispersion relation, vector mesons, 
generalized parton distributions}
\pacs{11.55.Fv, 13.40.Gp, 13.60.Hb, 13.66.Bc}
\preprint{NT@UW-10-20, JLAB-THY-10-1274}
\maketitle
\section{Introduction}
Learning to describe the structure and interaction of hadrons
on the basis of QCD is one of the main objectives of nuclear physics. 
An essential step in this program is to understand the structure
of the pion, a nearly massless excitation of the QCD vacuum with 
pseudoscalar quantum numbers. The pion plays a central role in 
nuclear physics as the carrier of the long--range force between 
nucleons and a harbinger of spontaneous symmetry breaking. 
The importance of the pion has been recognized by intense experimental 
and theoretical activity aimed at measuring its properties and 
understanding its structure. The pion electromagnetic form factor
$F_\pi(t)$ was measured at spacelike momentum transfers through
pion--electron scattering \cite{Dally:1981ur,Amendolia:1984nz} 
and pion electroproduction on the nucleon 
\cite{Bebek:1974ww,Ackermann:1977rp,Volmer:2000ek,Horn:2006tm}; 
new measurements in the region $|t| \sim \textrm{few GeV}^2$
are planned with the Jefferson Lab 12~GeV Upgrade \cite{E12-06-101}. 
In the timelike region the modulus of the (complex) pion form factor,
$|F_\pi(t)|$, was determined in a series of $e^+e^-$ experiments
\cite{Bollini:1975pn,Barkov:1985ac,Bisello:1988hq,%
Akhmetshin:2001ig,Pedlar:2005sj}; see Ref.~\cite{Whalley:2003qr} 
for a compilation of the older data.

The concept of transverse densities \cite{Soper:1976jc}, 
whose properties were explored in several recent works
\cite{Miller:2007uy,Miller:2009qu}, provides a model-independent 
way to relate the form factors of hadrons to their fundamental 
quark/gluon structure in QCD. Defined as the 2--dimensional 
Fourier transforms of the elastic form factors, the transverse densities
describe the distribution of charge and magnetization in the plane 
transverse to the direction of motion of a fast hadron; 
see Ref.~\cite{Miller:2010nz} for a review. 
They are closely related to the parton picture of hadron 
structure in high--energy processes and correspond to a reduction of 
the generalized parton distributions (or GPDs) describing the 
distribution of quarks/antiquarks with respect to longitudinal momentum 
and transverse position \cite{Burkardt:2000za,Diehl:2002he}.
It is therefore natural to attempt to interpret the pion form factor data 
in terms of the transverse charge density in the pion. In particular, 
the density at small transverse distances 
$b \ll 1\, \textrm{fm}$ places constraints
on the probability of pointlike configurations (or PLCs) in the 
pion --- $q\bar q$ configurations in the partonic wave function 
of a transverse size much smaller than the typical hadronic 
radius \cite{Frankfurt:1993es}.
Such configurations play an important role in high--momentum
transfer reactions involving pions, such as the pion transition
form factor $\gamma^\ast \gamma \rightarrow \pi^0$ 
\cite{Lepage:1980fj,Musatov:1997pu} or pion production in 
large--angle scattering processes \cite{Kumano:2009he}. 
They are essential for the physics of the color transparency 
phenomenon predicted by QCD \cite{Bertsch:1981py,Frankfurt:1993it},
which is studied in high--energy pion dissociation on nuclear 
targets \cite{Frankfurt:1999tq,Aitala:2000hc} and electromagnetic pion 
knockout \cite{:2007gqa,Miller:2010eh} and is closely related to 
the existence of factorization theorems for hard meson production processes.
The dynamical origin of PLCs --- whether they are generated through 
perturbative QCD interactions with large--size configurations or by
non-perturbative mechanisms, remains a subject of intense study.

The transverse charge density in the pion is defined as 
the 2--dimensional Fourier transform of the spacelike
pion form factor,
\begin{equation} 
\rho_\pi (b) \;\; = \;\; 
\int\limits_0^\infty \frac{dQ}{2\pi} \, 
Q \, J_0 (Qb) \; F_\pi(t = -Q^2) ,
\label{def}
\end{equation}
where $F_\pi$ is regarded as a function of the invariant 
momentum transfer $t$.
The function $\rho_\pi (b)$ gives the probability that charge is located at
a transverse separation $b$ from the transverse center of momentum,
with $\int d^2 b \, \rho_\pi (b) = 1$. The definition Eq.~(\ref{def})
may in principle be used to calculate the charge density directly 
from the spacelike form factor data. In the nucleon case,
where the spacelike form factors can be extracted directly from
the measured $eN$ elastic scattering cross section
and are known up to rather large momentum transfers, 
this approach has been quite successful; see Ref.~\cite{Venkat:2010by} 
for an assessment of the uncertainties. In the pion case 
the spacelike form factor at momentum transfers above 
$Q^2 > 0.25 \, \text{GeV}^2$ was extracted only indirectly in 
electroproduction experiments on the nucleon $N(e,e'\pi)N'$,
with substantial model dependence, and is known only poorly 
at higher $Q^2$, rendering such a program difficult. 
However, for the pion one has another avenue for evaluating 
the transverse density, based on a dispersion representation for the 
pion form factor. Noting that the singularities of 
$F_\pi(t)$ as an analytic function of $t$ are confined
to a cut along the positive real axis starting at $t = 4 m_\pi^2$,
the form factor can be expressed as \cite{Bjorken:1965zz}
\begin{equation} 
F_\pi(t) \;\; = \;\; \int\limits_{4m_\pi^2}^\infty \frac{dt'}{t' - t + i0} 
\; \frac{\textrm{Im}\, F_\pi (t')}{\pi}
\label{disp}.
\end{equation}
The asymptotic behavior expected from perturbative QCD, 
$F_\pi(t) \sim \alpha_s(t)/|t|$ for $t\rightarrow\infty$,
allows the use of an unsubtracted dispersion relation
\footnote{Use of a subtracted dispersion relation in Eq.~(\ref{def})
would lead to an expression for the transverse density which
differs from Eq.~(\ref{rel}) by a delta function $\delta^{(2)}(\bm{b})$.
Subtractions therefore have no influence on the dispersion result
for the transverse density at finite $b$.}. 
Substitution of Eq.~(\ref{disp}) in Eq.~(\ref{def})
leads to the result \cite{Strikman:2010pu}
\begin{equation}
\rho_\pi (b) \;\; = \;\; \int\limits_{4m_\pi^2}^\infty \frac{dt}{2\pi} 
\; K_0(\sqrt{t} b) \; \frac{\textrm{Im}\, F_\pi(t + i0)}{\pi} .
\label{rel}
\end{equation}
This representation of the charge density as a dispersion integral over 
the imaginary part (or spectral function) of the timelike pion form 
factor has an interesting ``filtering'' property. The exponential drop--off 
of the modified Bessel function $K_0$ at large arguments causes the 
integrand of Eq.~(\ref{rel}) to decrease exponentially at large $t$ and 
ensures that only values $\sqrt{t} \sim 1/b$ in the spectral function are 
effectively sampled at a given distance $b$. In the nucleon case the
timelike form factor is measurable only at $t > 4 m_N^2$
and Eq.~(\ref{rel}) is not useful for calculating the transverse
density from data (it is, however, very useful for theoretical 
analysis; for example, the chiral large--distance component of the 
nucleon charge density at $b\sim m_\pi^{-1}$ can be obtained from the
calculable strength of the two--pion cut in the nucleon form factor 
near threshold \cite{Strikman:2010pu}).
In the pion case the physical region for the timelike form factor
starts at $t = 4\, m_\pi^2$, covering the entire range of the
dispersion integral, and Eq.~(\ref{rel}) becomes a practical method 
for calculating the charge density at all values of $b$. 
High--quality $e^+e^-$ annihilation data exist for values of $t$ up 
to $\sim 1\, \textrm{GeV}^2$, so that we hope to be able to 
determine $\rho_\pi (b)$ accurately for values of $b$ at least 
down to values of $b\sim 1 \, \textrm{GeV}^{-1} = 0.2 \, \textrm{fm}$.

The imaginary part of the pion form factor $\textrm{Im}\, F_\pi (t)$ 
entering in the dispersion representation Eq.~(\ref{rel}) is not 
measured directly in annihilation experiments. The $e^+e^- \rightarrow 
\pi^+\pi^-$ cross section is proportional to $|F_\pi(t)|^2$, 
and model--dependent input is generally needed to determine the phase. 
In the region of the $\rho$ meson resonance this problem was studied 
extensively long ago and is under good theoretical control. 
The phase of the first higher resonance $\rho'$ is strongly 
constrained by the dispersion integrals (sum rules) for the pion 
charge and the measured charge radius. At larger values of $t$ arguments 
based on perturbative QCD and local duality provide some guidance.
Combined with the filtering property of the dispersion integral
Eq.~(\ref{rel}), these constraints strongly reduce the model
dependence in the transverse density at $b \gtrsim 0.1 \, \textrm{fm}$. 
Our estimates below show that the this way of constructing $\rho_\pi (b)$ 
gives substantially more accurate results than use of the spacelike 
pion form factor data alone.

In this article we calculate the transverse charge density in the pion
in the dispersion representation Eq.~(\ref{rel}) using an empirical 
parametrization of the timelike pion form factor based on $e^+e^-$
annihilation and spacelike form factor data \cite{Bruch:2004py}.
We find that the density is determined to an accuracy of $\sim 10\%$ 
at transverse distances $b \sim 0.1 \, \textrm{fm}$, and 
substantially better at larger values. We thus obtain a precise 
two--dimensional image of the fast--moving pion, which can be 
interpreted in terms of its partonic structure in QCD. 
In particular, the density exhibits a pronounced rise at small $b$, 
as was observed earlier --- although with much lower precision --- in an
analysis based on the spacelike pion form factor \cite{Miller:2009qu}.
Using experimental information on the quark density in the pion,
we argue that such singular behavior of the charge density cannot be 
explained by large--size, $x \rightarrow 1$ configurations 
in the pion's partonic wave function and must therefore be attributed 
to PLCs. Our result thus places constraints on the probability of 
PLCs in the pion, which can be probed in other high momentum--transfer 
processes involving pions.

The plan of this paper is as follows. In Sec.~\ref{sec:parametrization} 
we briefly describe the main features of the pion form factor in
the timelike region and the elements of the parametrization of 
Ref.~\cite{Bruch:2004py}.
In Sec.~\ref{sec:density} we calculate the transverse charge density
and investigate its uncertainties at small distances. The implications 
for the pion's partonic structure and the presence of PLCs are discussed 
in Sec.~\ref{sec:structure}. Section~\ref{sec:chiral} discusses the
possible role of chiral dynamics in the pion transverse density
at large distances. A summary and suggestions for further studies
are presented in Sec.~\ref{sec:summary}.
\section{Timelike form factor parametrization}
\label{sec:parametrization}
In the energy region $\sqrt{t} \lesssim 1 \, \textrm{GeV}$ the measured 
pion form factor $|F_\pi(t)|^2$ is dominated by the $\rho$ meson 
resonance, with clearly visible effects of $\rho$--$\omega$ mixing
(see Ref.~\cite{Bruch:2004py} for a summary of the data). Theoretical
support for $\rho$ dominance at the amplitude level comes from the
observation that the $2\pi$ channel accounts for most of the annihilation
cross section, which allows one to relate the pion form factor to the
$\pi\pi$ scattering amplitude via elastic unitarity. In this region
the form factor is successfully described by the Gounaris--Sakurai (GS)
amplitude \cite{Gounaris:1968mw}, which is derived from an effective 
range expansion of the $\pi\pi$ phase shift and has the correct 
analytic structure. The neglect of certain off--shell terms 
$\propto (t - m_\rho^2)$ in the GS amplitude leads to a
Breit--Wigner (BW) type parametrization with energy--dependent width;
this simplified form also describes the $|F_\pi(t)|^2$ data
in the region $\sqrt{t} \lesssim 1 \, \textrm{GeV}$ but does not 
respect the analytic properties of the form factor
(it has a spurious branch cut singularity at $t = 0$). 
We shall employ the full GS parametrization in our studies here.

Above the $\rho$ region data for $|F_\pi(t)|^2$ exist up to energies 
$\sqrt{t} \lesssim 3 \, \textrm{GeV}$. Because of the many hadronic
channels in the total cross section, the phase of the form factor at 
these energies is much more uncertain. In the region of the first 
higher resonance $\rho'$ the phase is constrained by the 
sum rules for the pion charge and the charge radius, which require
partial compensation of the spectral strength in the $\rho$ meson region. 
At higher energies theoretical constraints come from the
asymptotic behavior predicted by perturbative QCD, which demands
strong cancellations between higher resonances in a resonance--based 
description, as indeed found in dual resonance models. 

In the present study we use the timelike pion form factor parametrization
of Ref.~\cite{Bruch:2004py}, which describes the high--energy region 
by a pattern of resonances consistent with the QCD asymptotic behavior. 
The parameters were determined by a detailed analysis of the timelike 
data up to $\sqrt{t} \lesssim 3 \, \textrm{GeV}$. The continuation of these 
parametrizations to $t < 0$ also describes the spacelike form factor 
in accordance with the data, including the recent JLab data up to
$|t| = 2.45 \, \textrm{GeV}^2$ \cite{Horn:2006tm}, which appeared after 
publication of Ref.~\cite{Bruch:2004py}.

\begin{figure}
\includegraphics[width=0.48\textwidth]{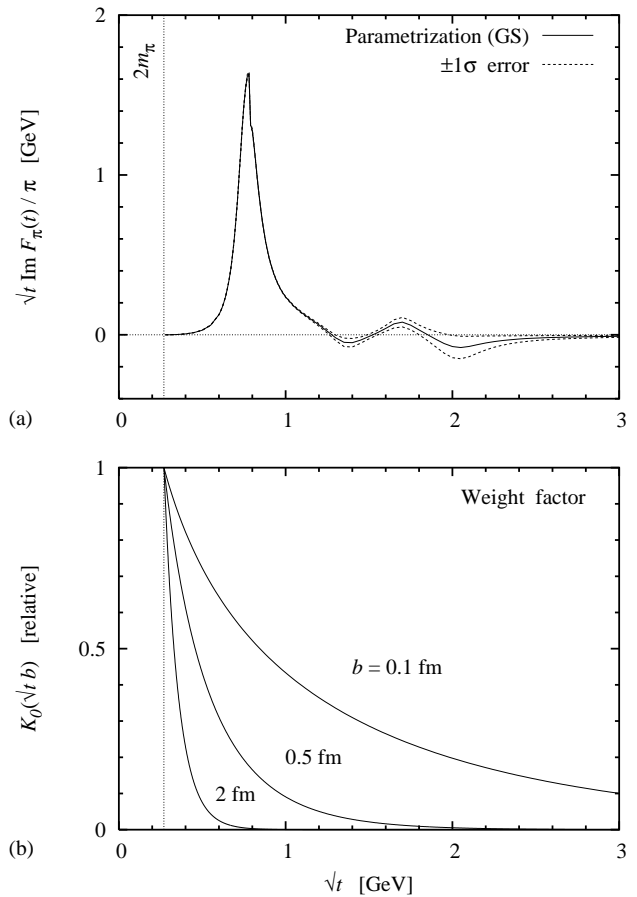}
\caption{(a) Solid line: The imaginary part of the pion form factor 
obtained from the fit of Ref.~\cite{Bruch:2004py} (GS parametrization), 
as a function of $\sqrt{t}$. Shown here is the function 
$\sqrt{t} \, \textrm{Im} F_\pi (t)/\pi$, which effectively enters in 
the dispersion integral over $\sqrt{t}$, Eq.~(\ref{rel}). 
Dotted lines: $\pm 1 \sigma$ error resulting from the 
uncorrelated uncertainties of the fit parameters.
The threshold energy $\sqrt{t} = 2 m_\pi$ is indicated by
the vertical line. 
(b) The weight factor $K_0 (\sqrt{t} b)$ in the dispersion
representation of the transverse charge density Eq.~(\ref{rel}), 
as a function of $\sqrt{t}$, for several values of $b$. Shown 
are the functions normalized to unity at the threshold $\sqrt{t} = 2 m_\pi$.}
\label{fig:fpi_im}
\end{figure}

A brief description of the elements of the parametrization 
of Ref.~\cite{Bruch:2004py} is provided here; for details we refer 
to the original article and references therein. The first four $\rho$
meson resonances are included as specific states with masses up to 
$2.0 \, \textrm{GeV}$ ($\rho$--$\omega$ mixing is taken into account 
for the lowest resonance). These resonances are described by the GS form, 
which incorporates the proper threshold behavior of the widths
and has the correct analytic 
properties \footnote{Reference~\cite{Bruch:2004py} also provides
an alternative fit in which the resonances are described by the
BW form. As mentioned above, the BW form has 
a spurious branch cut singularity at $t = 0$, which in the full GS form 
is canceled by the off-shell term. This has the consequence that 
the imaginary part obtained with the BW parameters 
of Ref.~\cite{Bruch:2004py} does not satisfy the normalization condition 
$\pi^{-1} \int_{4 m_\pi^2}^\infty dt' \, \textrm{Im}\, F_\pi (t')/t' = 1$
exactly, but with a small discrepancy related to the spectral 
strength on the unphysical cut (the BW ansatz in \cite{Bruch:2004py} 
was normalized to unit value at $t = 0$, not to unit integral over the
imaginary part). In our dispersion analysis we therefore
use the GS parametrization, which has the correct analyticity structure
and is free of this problem.}. In addition, an infinite series of 
higher excitations is included via an ansatz \cite{Dominguez:2001zu} 
based on the dual resonance model.
Its continuation to the space-like region exhibits a smooth behavior 
with a power--law asymptotics as $|t|^{1 - \beta}$ with $\beta = 2.1-2.3$.  
The imaginary part of the form factor obtained with 
the GS parametrization \cite{Bruch:2004py} is shown in 
Fig.~\ref{fig:fpi_im}a (solid line). One clearly sees the dominance
of the $\rho$ meson pole in the region $\sqrt{t} < 1 \, \textrm{GeV}$, 
and the alternating sign of successive resonance contributions
at larger values of $\sqrt{t}$, as expected from theoretical 
considerations.
%
%

To estimate the uncertainty in the imaginary part, we have taken 
the quoted variances of the fit parameters of Ref.~\cite{Bruch:2004py}
and studied the statistical variation of the imaginary part, assuming 
uncorrelated errors. The resulting $\pm 1\sigma$ error band is shown 
in Fig.~\ref{fig:fpi_im}a (dotted lines). The variance in the $\rho$ 
meson mass region is at the few percent level. At energies above
$1\, \textrm{GeV}$ it becomes substantially larger, reaching close to 
$100\%$ at $\sqrt{t} = 2 \, \textrm{GeV}$. Note that in this
energy region our uncorrelated estimate likely represents an 
upper bound on the uncertainty, as correlations between the statistical 
fluctuations of the coupling and width of the second resonance 
would considerably reduce the overall fluctuations of the 
imaginary part near $\sqrt{t} \sim 1.4 \, \textrm{GeV}$.
For energies above 3 GeV we cannot reliably estimate the relative
uncertainty of the imaginary part in this way, as the couplings of
the resonances in this region are dictated by the dual resonance model,
and the data in this region are very poor. However, the imaginary
part in this region is expected to be very small and contributes 
negligibly to the charge density at $b > 0.1 \, \textrm{fm}$ (see below),
so that its relative uncertainty is not important for our purposes.
We emphasize that we use the parametrization of Ref.~\cite{Bruch:2004py}
only as an effective representation of $\textrm{Im}\, F_\pi (t)$
in the energy range $\sqrt{t} < 3 \, \textrm{GeV}$, and that 
our conclusions do not depend on the particular 
$\sqrt{t} \rightarrow \infty$ asymptotic behavior 
imposed by the dual resonance model.
\section{Transverse density and its uncertainty}
\label{sec:density}
We now use the timelike form factor parametrization
to evaluate the transverse charge density in the pion and estimate
its uncertainty. It is instructive 
to study first the distribution of strength in the dispersion integral 
Eq.~(\ref{rel}). The imaginary part $\textrm{Im}\, F_\pi (t)$ is weighted 
with the modified Bessel function $K_0 (\sqrt{t} b)$, which exponentially 
suppresses energies $\sqrt{t} \gg 1/b$. Fig.~\ref{fig:fpi_im}b
shows this weight factor as a function of $\sqrt{t}$ for several
values of $b$, normalized to the same value at threshold 
$\sqrt{t} = 2 m_\pi$, i.e., the ratio
\begin{equation}
K_0 (\sqrt{t} b) / K_0 (2 m_\pi b) .
\end{equation}
One sees that the effective distribution of strength in $\sqrt{t}$
strongly changes with the distance $b$. At $b = 0.1 \, \textrm{fm}$ 
a noticeable contribution to the dispersion integral 
comes from the region $\sqrt{t} > 1 \, \textrm{GeV}$, where the 
parametrization of $\textrm{Im}\, F_\pi (t)$ shows considerable 
uncertainty (see Fig.~\ref{fig:fpi_im}a). At $b = 0.5 \, \textrm{fm}$ 
these contributions are largely suppressed, resulting in almost 
perfect ``vector meson dominance'' in the dispersion integral.
Finally, going to distances as large as $b \sim 2 \, \textrm{fm}$, 
one begins to suppress also the $\rho$ mass region and emphasizes 
the near--threshold region of the form factor, 
$\sqrt{t} - 2 m_\pi \sim \textrm{few} \, m_\pi$.

\begin{figure}
\includegraphics[width=0.48\textwidth]{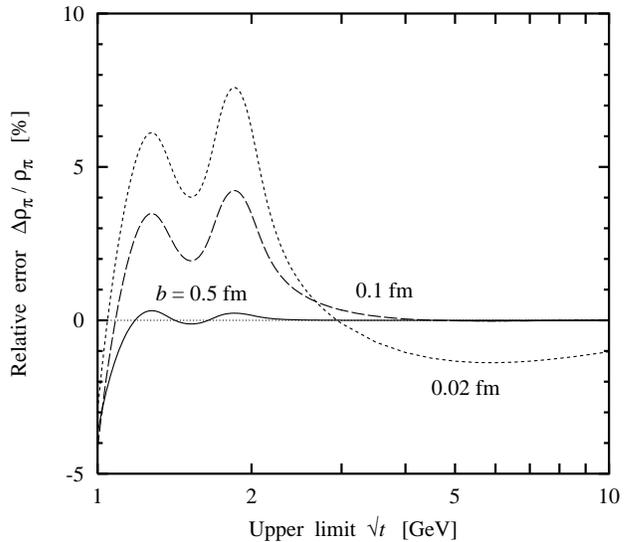}
\caption{Percentage deviation from the full result 
for the dispersion integral Eq.~(\ref{rel}), as a function of the 
upper limit of $\sqrt{t}$, for $b = 0.5 \, \textrm{fm}$ (solid line),
$0.1 \, \textrm{fm}$ (dashed line) and $0.02 \, \textrm{fm}$ (dotted line). 
The integrand is evaluated using the 
GS parametrization of Ref.~\cite{Bruch:2004py}.}
\label{fig:error}
\end{figure}

In order to quantify the accuracy of the calculated transverse density 
we need to study the numerical convergence of 
the dispersion integral at large values of $\sqrt{t}$. 
Figure~\ref{fig:error} shows the percentage deviation of $\rho_\pi (b)$ from 
the full result as a function of a cutoff applied to the upper limit 
of the $\sqrt{t}$ integral in Eq.~(\ref{rel}) (here the integral is
evaluated with the central value of the GS parametrization 
as shown in Fig.~\ref{fig:fpi_im}). One sees that 
at $b = 0.1 \, \textrm{fm}$ the region $\sqrt{t} > 3 \, \textrm{GeV}$
accounts for only about $\sim 1\, \%$ of the total integral, meaning
that even a drastic change of $\textrm{Im}\, F_\pi (t)$ in this region
by a factor $2-3$ would change the density only by $\sim 2-3\%$
\footnote{This argument assumes that the change of 
$\textrm{Im}\, F_\pi (t)$ does not substantially alter the cancellation 
of successive resonance contributions implied by the dual resonance
picture. The effect of an ``unbalanced'' resonance will be 
estimated separately below.}. The error in the density is thus
dominated by the mass region $1 < \sqrt{t} < 3 \, \textrm{GeV}$, 
where we have estimated the uncertainty of $\textrm{Im}\, F_\pi (t)$
in Sec.~\ref{sec:parametrization}. With a $\sim 100\%$ uncertainty at
$\sqrt{t} = 2 \, \textrm{GeV}$, where the integral has converged
to within $\sim 4\%$ of its value, we expect an uncertainty of
the density of (at least) $\sim 4\%$. Surprisingly, even for 
much smaller distances the region $\sqrt{t} > 3 \, \textrm{GeV}$
seems to contribute relatively little to the dispersion integral;
see the curve in Fig.~\ref{fig:error} for $b \sim 0.02 \, \textrm{fm}$.
While the integral requires larger values of $\sqrt{t}$ to converge,
the contribution from $\sqrt{t} > 3 \, \textrm{GeV}$ is still 
only $\sim 2\%$, and the overall uncertainty can be estimated from that 
of the $1 < \sqrt{t} < 3 \, \textrm{GeV}$. At larger distances 
$b\sim 0.5 \, \textrm{fm}$, however, the integral has fully converged
already at $\sqrt{t} \sim 1\, \textrm{GeV}$, and the overall
uncertainty is dominated by the low--energy region
$\sqrt{t} < 1\, \textrm{GeV}$. In this region the parameter errors
in the fit are so small that the model dependence of the parametrizations
(details of $\rho$--$\omega$ mixing, $\rho$ line shape) can no
longer be neglected in establishing the overall error.
%
%

\begin{figure}
\includegraphics[width=0.46\textwidth]{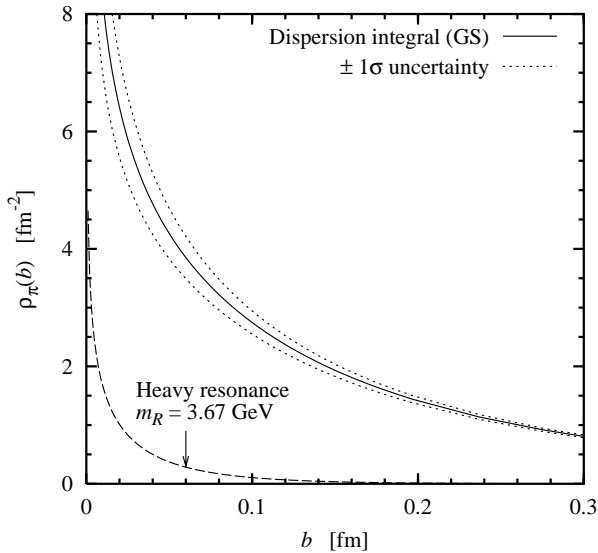} 
\caption{Transverse charge density in the pion, $\rho_\pi (b)$. 
Solid line: Dispersion integral Eq.(\ref{rel})
evaluated with the GS form factor parametrization \cite{Bruch:2004py}
(see Fig.~\ref{fig:fpi_im}a). 
Dashed lines: $1\sigma$ error resulting from the quoted
uncertainty of the parametrization (see Fig.~\ref{fig:fpi_im}a).
Dotted line: Density resulting from a heavy resonance
with mass $m_R = 3.67 \, \textrm{GeV}$ and width $\Gamma_R = 0.2 \, m_R$,
providing a rough assessment of the impact of the CLEO timelike 
form factor data \cite{Pedlar:2005sj} (details see text).}
\label{fig:rho_lin}
\end{figure}

Given the dominance of energies $\sqrt{t} < 3 \, \textrm{GeV}$
in the dispersion integral, we can evaluate the density with the
parametrization of Ref.~\cite{Bruch:2004py} and estimate its
uncertainty from the parameter error band shown in Fig.~\ref{fig:fpi_im}
The result is displayed in Fig.~\ref{fig:rho_lin}. The quoted $1\sigma$ 
error in $\textrm{Im}\, F_\pi (t)$ translates into an uncertainty
of $\rho_\pi (b)$ of $\pm (1.5, \, 7, \, 13)\%$ 
at $b = (0.5, \, 0.1, \, 0.02) \, \textrm{fm}$.
The density is thus determined much more accurately, and down to much 
smaller distances, than from the spacelike pion form factor 
data alone \cite{Miller:2009qu}.

A welcome feature of the dispersion representation of the 
charge density, Eq.~(\ref{rel}), is that the kernel $K_0 (\sqrt{t} b)$
is a positive function. As a result, an upper or lower bound on the 
spectral function $\textrm{Im}\, F_\pi (t)$ directly provide a 
corresponding bound on $\rho_\pi (b)$, greatly simplifying the 
error analysis. (A method to estimate the uncertainty of the 
charge density as the Fourier transform of the spacelike form
factor was described in Ref.~\cite{Venkat:2010by}.)

An additional source of uncertainty in the charge density at small
distances are recent data on the timelike pion form factor at large 
values of $t$ that were not included in the fit of Ref.~\cite{Bruch:2004py}. 
The CLEO measurement \cite{Pedlar:2005sj} at 
$\sqrt{t} = 3.67\, \textrm{GeV}$ reports a value of 
$|F_\pi| = 0.075 \pm 0.008\, \textrm{(stat)} \pm 0.005 \textrm{(syst)}$,
much larger than the value 0.034 provided by the GS parametrization of
Ref.~\cite{Bruch:2004py}. We see no simple 
way to modify the parametrization to account for this datum. 
Indeed, Ref.~\cite{Bruch:2004py} argues that increasing the
absolute value of the form factor by a factor of $\sim 2$ at large
$\sqrt{t}$ is not possible. In particular, the article states that 
it is implausible for the form factor obtained on the basis of a
dual resonance parametrization to reach values $|F_\pi(t)|^2 \ge 0.01$ 
at $\sqrt{t} = 2.5-3\, \textrm{GeV}$ (as would correspond to the
new datum, assuming power-like $t$--dependence) without conflicting 
with the spacelike data and especially with QCD predictions
\footnote{We note that the fit of Ref.~\cite{Bruch:2004py}
also does not reproduce the value of $|F_\pi|^2$ at 
$\sqrt{t} = 3.1\, \textrm{GeV}$ extracted from the $J/\psi \rightarrow
\pi\pi$ decay; see Ref.~\cite{Bruch:2004py} for a critical discussion
of this datum.}.
%
%

One possibility is that the error of the CLEO result is 
larger than estimated in Ref.~\cite{Pedlar:2005sj}. Another possibility 
is that there is a new mechanism providing a strong coupling to
two pions at high energies. Here we only wish to make a rough assessment 
of the potential impact of this new datum on the transverse density. 
To this end, let us assume the existence of an ``additional'' 
$\pi\pi$ resonance at $\sqrt{t} = m_R = 3.67\, \textrm{GeV}$, 
described by the GS form, whose coupling $c_R$ to the virtual photon 
is related to the measured pion form factor as
\begin{equation}
|F_\pi (m_R)| \;\; = \;\; c_R m_R / \Gamma_R .
\end{equation}
Taking the width $\Gamma_R$ to be
$\sim 20\%$ of the mass, as it is for the $\rho$ meson,
we obtain a coupling $c_R = 0.015$ from the CLEO measurement. 
Such an addition gives a 
negligible contribution to the $|F_\pi(t)|$
at values of $\sqrt{t}$ for which most of the data entering
in the parametrization \cite{Bruch:2004py} were taken;
for example, it provides a  $\sim 1\%$ contribution 
to $|F_\pi(t)|$ at $\sqrt{t} = 1\, \textrm{GeV}$.
The ``extra'' contribution to the charge density from such a
resonance would be $+(0.04, \, 4, \, 16)\, \%$ at 
$b = (0.5, \, 0.1, \, 0.02) \, \textrm{fm}$ (see Fig.~\ref{fig:rho_lin}). 
If we added this uncertainty to the one estimated 
previously from the error of the parametrization for 
$\sqrt{t} < 1 \, \textrm{GeV}$, we would conclude that the density 
is determined to $(\pm 1.5, +11-7, +39-16)\, \%$ at the quoted values of $b$.
This is surely a conservative estimate, as at least part of 
the uncertainty in the unmeasured high--$t$ region is already 
included in the parametrization error. A larger value of the width of 
the hypothetical resonance would lead to a proportionately larger 
contribution to $\rho_\pi (b)$, but would have to be reconciled with 
the precise data for $|F_\pi|$ in the mass 
region $\sqrt{t} \lesssim 1 \, \textrm{GeV}$. 
We conclude that the new CLEO data have only a modest impact
on the transverse density at distances $\sim 0.1\, \textrm{fm}$, 
but may cause substantial modifications at smaller distances. 

%
%
\begin{figure}
\includegraphics[width=0.48\textwidth]{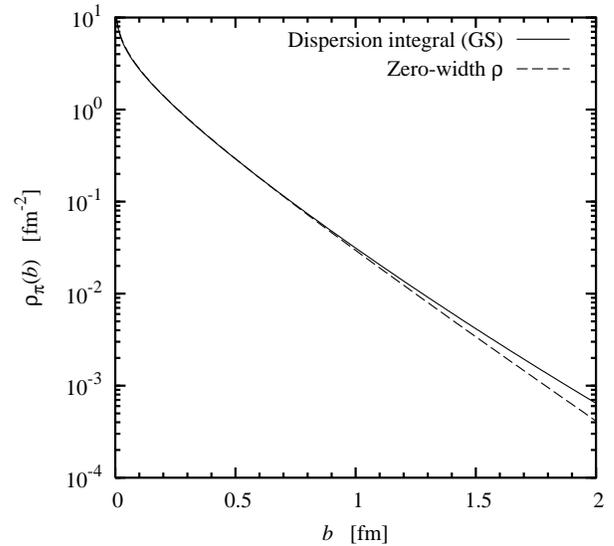} 
\caption{Transverse charge density $\rho_\pi (b)$ in the 
pion. Solid line: Dispersion integral Eq.(\ref{rel})
evaluated with the GS form factor parametrization of 
Ref.~\cite{Bruch:2004py}. Dashed line: Density
from zero--width $\rho$ meson pole, Eq.~(\ref{rho_pole}).}
\label{fig:rho_log}
\end{figure}
Figure~\ref{fig:rho_log} shows the transverse density obtained from the 
dispersion integral on a logarithmic scale, which allows one
to see the exponential fall--off at larger distances.
For comparison we also show the density
obtained from a single resonance of zero width at the $\rho$ meson mass 
$m_\rho$, with a coupling chosen to ensure unit charge 
(i.e., the vector meson dominance model)
\begin{equation}
\rho_\pi (b)_{\rm zero-width} \;\; = \;\; (m_\rho^2 / 2\pi) \, K_0(m_\rho b).
\label{rho_pole}
\end{equation}
One sees that the dispersion result is very close to the zero--width $\rho$
form for all distances $0.1 < b < 1 \, \textrm{fm}$ and can be represented
by the latter within the estimated errors (at larger values of $b$
the spectral strength near threshold becomes important; see 
Sec.~\ref{sec:chiral}). What is more, the dispersion result follows 
the zero--width $\rho$ curve down to much smaller distances, 
being only a few percent smaller down to $b = 0.01 \, \textrm{fm}$. 
This shows that there are very strong cancellations between the effective 
poles parametrizing the high--mass continuum. As we just demonstrated,
there is considerable uncertainty in the dispersion result for the 
density at such small distances. However, there is the intriguing 
possibility that the density might effectively be described by 
vector meson dominance down to distances significantly smaller than 
the inverse $\rho$ meson mass, $m_\rho^{-1} = 0.25\, \textrm{fm}$.

%
%
\begin{figure}
\includegraphics[width=0.48\textwidth]{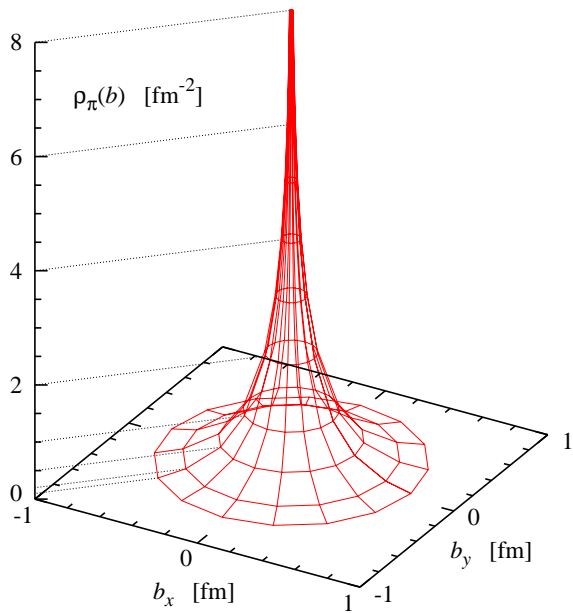} 
\caption{(Color online) Three--dimensional rendering of the transverse 
charge density in the pion, as obtained from the dispersion integral
Eq.(\ref{rel}) evaluated with the GS form factor parametrization of 
Ref.~\cite{Bruch:2004py}; cf.\ Figs.~\ref{fig:rho_lin}
and \ref{fig:rho_log}.}
\label{fig:rho3d}
\end{figure}
In Fig.~\ref{fig:rho3d} we show a 3--dimensional rendering of the
transverse charge density, which conveys also the information on the 
supporting area and thus gives an impression of the true physical
shape of the fast--moving pion as seen by an electromagnetic probe. 
Our dispersion approach provides a data--based image of the pion's 
transverse structure at small distances with unprecedented precision.
One clearly sees the strong rise of the transverse density toward
the center. This remarkable observation calls for a microscopic 
explanation in terms of the pion's partonic structure.
\section{Implications for pion partonic structure}
\label{sec:structure}
\begin{figure}
\includegraphics[width=0.48\textwidth]{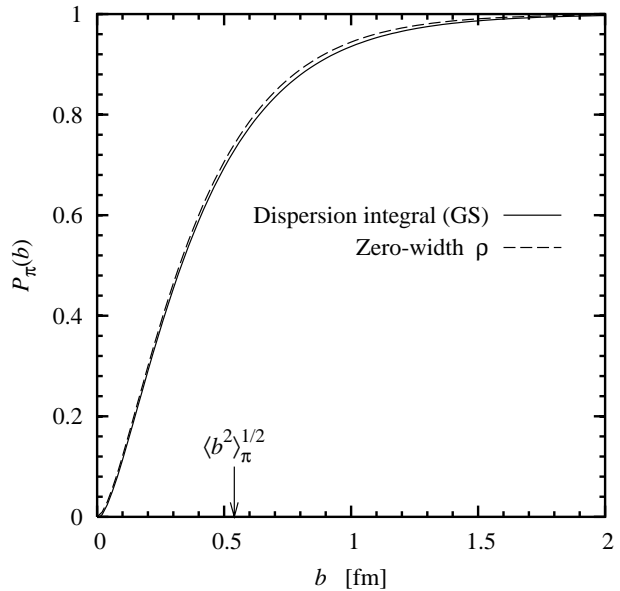} 
\caption{Probability accumulation Eq.~(\ref{pba}) in the transverse 
density (cf.\ Figs.~\ref{fig:rho_lin} and \ref{fig:rho_log}). 
Solid line: Dispersion integral (GS parametrization). 
Dashed line: Zero--width $\rho$ meson pole.
The arrow indicates the experimental RMS transverse charge radius.}
\label{fig:prob}
\end{figure}

The results of our empirical study of the transverse charge density
have interesting implications for the partonic structure of the pion
in QCD. The transverse charge density puts constraints on the possible
distribution of transverse sizes of configurations in the pion's 
partonic wave function. A useful quantity to consider is the 
integral of the transverse charge density up to a given distance,
\begin{equation}
P(b) \;\; \equiv \;\; \int d^2b \; \Theta(b - b') \; \rho_\pi (b'),
\label{pba}
\end{equation}
which determines the cumulative probability for configurations 
contributing to the transverse density at the distance $b$.
The probability obtained from our dispersion result for the charge
density (cf.\ Figs.~\ref{fig:rho_lin} and \ref{fig:rho_log}) 
is shown in Fig.~\ref{fig:prob},
together with that obtained from a zero--width $\rho$ 
meson pole (cf.\ Eq.~\ref{rho_pole}),
\begin{equation}
P(b)_{\rm zero-width} \;\; = \;\; m_\rho b \, K_1(m_\rho b) .
\label{prob_pole}
\end{equation}
The probability reaches $1/2$ at $b = 0.33 \, \textrm{fm}$, a value 
somewhat smaller than the root of the mean squared (RMS) transverse 
radius, $\langle b^2 \rangle^{1/2}_\pi = 0.53 \, \textrm{fm}$.
This is to be expected, as large--size configurations are counted
with a higher weight in the average of $b^2$ than than the median.
The RMS transverse radius calculated from our dispersion integral
for the charge density agrees very well with the value extracted 
from the slope of the low--$t$ pion form factor measured in $\pi e$ 
scattering experiments, 
$\langle r^2 \rangle_\pi = (3/2) \langle b^2 \rangle_\pi
= 0.439 \pm 0.008 \, \textrm{fm}^2$ \cite{Dally:1981ur,Amendolia:1984nz},
as was already noted in the discussion of the fit to the timelike 
form factor data in Ref.~\cite{Bruch:2004py}.
%
%

To understand how the transverse charge density is related to the
partonic structure it is necessary to recall the relationship between 
the coordinate $b$ and the physical transverse size of configurations 
in the fast--moving pion. The coordinate $b$ measures
the distance between a constituent --- say, a quark $q$ --- and the 
transverse center of mass of the pion. If the quark carries longitudinal 
momentum fraction $x$, and the remnant system $R$ carries $1 - x$, 
the transverse center of momentum of the pion is at 
$x \bm{r}_{q} + (1 - x) \bm{r}_R$, where $\bm{r}_{q, R}$ denotes 
the transverse position of the quark and the center of momentum 
of the remnant system. The transverse separation of the quark
from the remnant system is thus given by
\begin{equation} 
r \;\; \equiv \;\; |\bm{r}_q - \bm{r}_{R}| \;\; = \;\; b/(1 - x).
\label{r_b_relation}
\end{equation}
Figure~\ref{fig:impact} illustrates this relation for a $q\bar q$
configuration in which the remnant system consists of a single antiquark.
In the transverse charge density one considers the charge--weighted 
density of constituents at a given $b$, which is obtained as the average
over configurations with different $x$ and physical size $r$ in the 
partonic wave function. Equation~(\ref{r_b_relation}) now implies 
that the charge density at $b$ much smaller than the typical hadronic
size $R_{\rm had} \sim 1 \, \textrm{fm}$ can arise from two different 
classes of configurations:  
%
%
\begin{figure}
\includegraphics[width=0.27\textwidth]{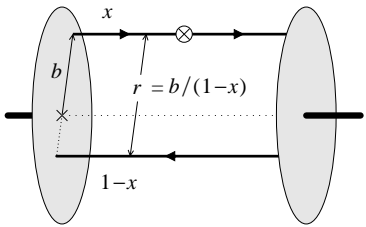} 
\caption{Transverse distances in a $q\bar q$ configuration of the
pion's partonic wave function: $b$ is the distance between the quark 
and the transverse center of momentum, $r$ the distance between the 
$q$ and $\bar q$.}
\label{fig:impact}
\end{figure}
\begin{enumerate}
\item[I)] Small physical size $r \ll R_{\rm had}$ and non--exceptional
values of $x$, i.e., not close to 1 (PLCs). 
One expects the elementary $q\bar q$ configuration accounts for a 
significant fraction of these configurations, as the emission of gluons 
and creation of additional $q\bar q$ pairs are suppressed in small--size 
color--singlet configurations. The existence of such PLCs is
required by the non--zero value of the pion weak decay matrix element,
parametrized by the constant $f_\pi$, where the axial current operator 
annihilates a $q\bar q$ pair in a point in space \cite{Lepage:1980fj}. 
\item[II)] Large physical size $r \sim R_{\rm had}$ and extreme 
momentum fractions $x \sim 1 - b/R_{\rm had}$ (end--point configurations).
These are generally not just $q\bar q$ configurations, as soft gluon
radiation is not suppressed in large--size configurations.
The probability of such configurations determines the behavior of the
parton densities in the pion at large values of $x$.
\end{enumerate}
As all analysis of partonic structure, the distinction between the 
two classes of configurations depends on the resolution scale $Q^2$.
Standard leading--$\log Q^2$ evolution degrades the parton momentum
fractions and reduces the probability of end--point configurations.
The total charge density resulting from the sum of all configurations
is of course scale--independent, being the matrix element of a
conserved current.

\begin{figure}
\includegraphics[width=0.48\textwidth]{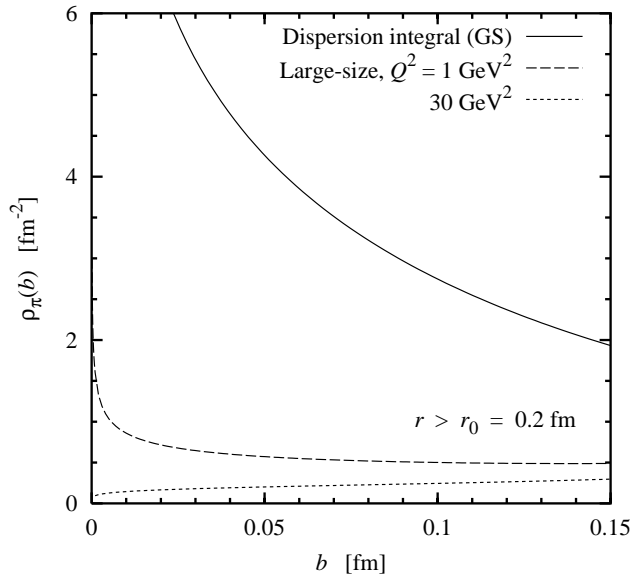}
\caption{Dashed/dotted line: Contribution of large--size configurations
with $r > r_0$ to the transverse charge density in the pion, 
as estimated in the model defined by
Eqs.~(\ref{rho_x_b})--(\ref{rho_b}), for $r_0 = 0.2$ and two
values of $Q^2$. Solid line: Density obtained from the 
dispersion integral, cf.\ Fig.~\ref{fig:rho_lin}.}
\label{fig:larges}
\end{figure}

We can estimate the possible contribution of large--size $x \rightarrow 1$ 
configurations to the transverse density at small $b$ in a
simple phenomenological model, using information on the quark 
distribution in the pion at large $x$ extracted from fits to 
$\pi N$ Drell--Yan data \cite{Conway:1989fs}. 
Our basic assumption here is that the physical
transverse size of large--$x$ configurations in the pion tends to 
a finite value of the order of the typical hadronic size.
Generalizing the expression obtained from the overlap of light--cone 
wave functions of individual configurations,
we model the $x$-- and $b$--dependent charge density 
(i.e., the charge--weighted quark GPD) arising from large--$x$
configurations as
\begin{equation}
\rho_\pi (x, b)_{\rm large-size} 
\;\; = \;\; q_\pi (x) \; \frac{f(r = b/(1-x))}{(1 - x)^2} ,
\label{rho_x_b}
\end{equation}
where $q_\pi (x)$ is the valence quark distribution in the pion 
\footnote{The valence quark density in Eq.~(\ref{rho_x_b})
is assumed to be normalized to unit integral, $\int dx \, q_\pi (x) = 1$.
Note, however, that Eq.~(\ref{rho_x_b}) is intended as a model
for large--$x$ configurations and should be integrated
only over this region.} and $f(r)$ describes the distribution 
over physical transverse sizes $r$, with a range of the order of the
typical hadronic size, normalized such that $\int d^2 r \, f(r) = 1$;
Eq.~(\ref{rho_x_b}) thus satisfies $\int d^2 b \, \rho_\pi (x, b) = q_\pi (x)$.
The transverse charge density $\rho_\pi (b)$ arising from large--size
configurations is then given by the integral of the density 
Eq.~(\ref{rho_x_b}) over $x$. In calculating this integral we impose 
the \textit{physical} requirement that the transverse size $r$ of the 
configuration be larger than some critical $r_0$.
This limits the range of $x$ in the integral to values
$x > 1 - b/r_0$, where it is assumed that $b < r_0$. We thus consider the 
``conditional'' large--size contribution to the density defined as
\begin{equation}
\rho_\pi (b| r > r_0) \;\; = \;\; \int_{1 - b/r_0}^1 
dx \, \rho_\pi (x, b)_{\rm large-size} .
\label{rho_b}
\end{equation}
To evaluate this contribution to the charge density at small $b$,
we use the the parametrization of the pion quark density of 
Ref.~\cite{Gluck:1999xe}. The size distribution $f(r)$ we take to 
be of Gaussian form, $f(r) =  \exp(-r^2/R^2)/(\pi R^2)$, where 
the parameter $R^2 = \langle r^2 \rangle_\pi$ defines the average 
squared radius and is of the order of the typical hadronic size 
$\sim 1 \, \textrm{fm}^2$. For a loosely bound $q\bar q$ state with 
$\langle x \rangle_\pi = 1/2$ one would have 
$\langle r^2 \rangle_\pi = \langle b^2/(1 - x)^2 \rangle_\pi 
\approx 4 \langle b^2 \rangle_\pi$; a natural choice is therefore 
$R^2 = 4 \langle b^2 \rangle_{\pi, {\rm exp}} = 1.16 \, \textrm{fm}^2$.
Figure~\ref{fig:larges} shows the contribution to the charge density 
from configurations with $r > r_0 = 0.2 \, \textrm{fm}$ estimated with 
this model, for two values of $Q^2$. One sees that it accounts only 
for at most $\sim 20\%$ of the total transverse density at 
$b = 0.1 \, \textrm{fm}$, and even less at smaller distances.
We thus conclude that large--size configurations with $x \rightarrow 1$
play only a minor role in the pion transverse charge density at small $b$,
and that most of it can be attributed to PLCs.

The small--$b$ behavior of the large--size contribution
to the charge density in our model can formally be related to the 
power behavior of the quark distribution in the pion for $x \rightarrow 1$.
In the limit of small $b$ the integral in Eq.~(\ref{rho_b}) 
extends over a narrow range of $x$ close to 1.
If the quark distribution vanishes as $q_\pi (x) \sim (1 - x)^\beta$, 
one easily shows that the density Eq.~(\ref{rho_b}) scales as
$\rho_\pi (b| r > r_0) \sim b^{\beta - 1}$ for $b \rightarrow 0$.
The change in the small--$b$ behavior with $Q^2$ seen in 
Fig.~\ref{fig:larges} reflects the effect of QCD evolution 
on the exponent $\beta$. Note, however, that even in the low--$Q^2$
region where $\beta < 1$ the large--size contribution in our model
is substantially smaller than the total density obtained from the
dispersion integral.
%
%

In sum, our estimate shows that large--size $x \rightarrow 1$ 
configurations cannot account 
for the strong rise of the transverse density at small $b$, and that it 
is therefore reasonable to interpret the empirical density in terms of 
PLCs in the pion's partonic wave function. In $q\bar q$ configurations 
of small size, it is expected that the
wave function peaks at $x = 1/2$, which implies that the physical
transverse size of the most likely configurations is $r \approx 2 b$. 
With the plausible assumption that the small--size configurations 
in the pion are mostly $q\bar q$, 
we would conclude from Fig.~\ref{fig:prob} that there is a 
probability of 12\% (29\%) for configurations with 
$b < 0.1 \, \textrm{fm} \; (0.2 \, \textrm{fm})$, and thus with $q\bar q$
separation $r \lesssim 0.2 \, \textrm{fm} \; (0.4 \, \textrm{fm})$.
In reality, some of these small--size configurations are 
$q\bar q + \textrm{gluons}$ or $qq\bar q\bar q$, requiring a detailed 
model--dependent analysis. Even so, our result for the charge density 
places strong constraints on the pion's partonic structure at small 
distances. The study of dynamical models of PLCs in the pion and their 
comparison with the empirical charge density will be the subject of 
future work.
\section{Long--range pion structure and chiral dynamics}
\label{sec:chiral}
To complete our study of the empirical transverse charge density 
in the pion we briefly want to comment on the possible role of chiral
dynamics at large transverse distances. 
At $b \gtrsim m_\pi^{-1} = 1.5 \, \textrm{fm}$ the weighting factor 
$K_0 (\sqrt{t} b)$ in the dispersion integral Eq.~(\ref{rel})
emphasizes the near--threshold region $\sqrt{t} - 2 m_\pi \sim 
\textrm{few} \; m_\pi$ (see Fig.~\ref{fig:fpi_im}b), 
where the imaginary part of the form factor 
is governed by chiral dynamics and calculable from first principles
In leading order of the chiral expansion, the imaginary
part near threshold results from the pion loop graph with the 
$\pi\pi$ 4--point coupling and is given 
by \cite{Beg:1973sc,Gasser:1983yg,Gasser:1990bv}
\begin{equation}
\pi^{-1} \, \textrm{Im} \, F_\pi (t + i0) \;\; = \;\;
\frac{(t - 4 m_\pi^2)^{3/2}}{6 (4\pi f_\pi)^2 \sqrt{t}} .
\label{chiral}
\end{equation}
Substitution of this result in Eq.~(\ref{rel}) allows one to
derive the leading $\exp(-2 m_\pi b)$ asymptotic behavior 
of the pion charge density at large distances; see 
Ref.~\cite{Strikman:2010pu} for details. Numerical analysis shows that 
the contribution from Eq.~(\ref{chiral}) to the charge density is 
negligible compared to the non--chiral density resulting from 
$\sqrt{t} \sim m_\rho$ for all but the largest distances, 
reaching only $\sim 30\%$ of the dispersion result at $b = 2 \, \textrm{fm}$.
In the nucleon isovector form factor the chiral component of the charge 
density was found to become comparable to the non--chiral density at
distances $b \sim 1.7 \, \textrm{fm}$ \cite{Strikman:2010pu};
the reason for its diminished role in the pion charge density 
is that the triangle graph involving the $\pi N$ Yukawa 
coupling (see Fig.~1 of \cite{Strikman:2010pu}),
which gave the main contribution in the nucleon case, is absent for the pion.
Account of higher--order chiral corrections does not substantially
change the magnitude of the chiral component \cite{Gasser:1990bv}. 
We conclude that the transverse charge density in the
pion is dominated by the $\rho$ meson mass region for all 
distances of practical relevance, $b < 2 \, \textrm{fm}$.
\section{Summary and discussion}
\label{sec:summary}
This paper shows how the pion form factor in the timelike region can be 
used to determine the transverse charge density. The timelike data 
greatly augment the meager information available from spacelike 
pion form factor measurements, in particular in the region of high
momentum transfers $|t| > 1\, \textrm{GeV}^2$ conjugate to short
transverse distances. Given the energy reach of the timelike form
factor data, and the theoretical uncertainties involved in separating
the real and imaginary parts, we estimate that $\rho_\pi (b)$ is determined
to an accuracy of $\sim 10\%$ at $b = 0.1 \, \textrm{fm}$, and
substantially better at larger distances. The transverse density obtained 
from the full dispersion integral turns out to be surprisingly close to 
that obtained from a single zero--width $\rho$ meson pole over a wide
range. The empirical transverse density shows a strong rise at small
distances, which points to a substantial presence of PLCs in the
pion's partonic wave function and puts strong constraints on the 
pion GPD.

In the work reported here we limited ourselves to a phenomenological
analysis of the transverse density based on an existing parametrization 
of the timelike pion form factor data. Our results suggest several 
directions for further studies, both theoretical and empirical.

The striking similarity of the empirical transverse density to the
simple vector meson dominance model over a wide region of $b$ should have
a dynamical explanation. Possible approaches to address this
question are local quark--hadron duality or the dual resonance picture 
of QCD in the large--$N_c$ limit.

The strong rise of the pion's transverse charge density at small
distances calls for an explanation in terms of dynamical models
of the pion's partonic structure. The key question is whether the
required PLCs in the pion could be explained as the result of
perturbative QCD interactions with large--size configurations,
or whether non-perturbative interactions play an essential role.
Of particular interest for addressing this question are models
which implement the non--perturbative short--distance scale associated 
with the spontaneous breaking of chiral symmetry in QCD such as
the instanton vacuum model, which is known to give a reasonable description 
of the spacelike pion form factor at intermediate momentum transfers 
$Q^2 \sim \textrm{few GeV}^2$ \cite{Diakonov:1985eg,Faccioli:2002jd}.

The dispersion result for the transverse charge density at distances 
$b \sim 0.1 \, \textrm{fm}$ depends sensitively on the phase of the 
pion form factor in the region of the lowest excited $\rho$ states,
$\sqrt{t} = 1-3 \, \textrm{GeV}$. While the alternating sign of the
coupling of successive resonances is suggested by theoretical
considerations, it would be worthwhile to attempt independent 
experimental tests of this key assumption. This could be done 
through coherent photo-- or electroproduction of two pions on 
nuclear targets, which can be analyzed in the spirit of the 
generalized vector meson dominance model; 
see Ref.~\cite{Frankfurt:2003wv} and references therein. Such 
measurements become feasible with the 12~GeV Upgrade of Jefferson Lab.

The recent CLEO data \cite{Pedlar:2005sj}, which are difficult to 
explain in the dual resonance framework commonly used to parametrize the 
high--energy region of the pion form factor, may have a significant 
effect on the charge density at distances $b < 0.1 \, \textrm{fm}$.
Confirmation of this experimental result and more data in the energy
region $\sqrt{t} = 3 - 4 \, \textrm{GeV}$ would certainly be welcome.
It would be interesting to explore ways to include these data in a
dispersion analysis with more general parametrizations of the 
imaginary part.

The new application of the timelike pion form factor described here
once more underscores the importance of analyticity in relating
observables measured in different kinematic regions. It would be
helpful if phenomenological parametrizations of the form factors 
such as \cite{Bruch:2004py} employed a framework which strictly 
respects analyticity, e.g.\ by using analytic functions
like the GS form, or by parametrizing only the spectral strength 
on the physical cut and generating the real part by a 
dispersion integral.

\section*{Acknowledgments}
G.~A.~M.\ acknowledges the hospitality of Jefferson Lab during the work 
on this study. This work is supported by the U.S.\ DOE under Grants No. 
DE-FGO2-97ER41014 and DE-FGO2-93ER40771.
Notice: Authored by Jefferson Science Associates, LLC under U.S.\ DOE
Contract No.~DE-AC05-06OR23177. The U.S.\ Government retains a
non--exclusive, paid--up, irrevocable, world--wide license to publish or
reproduce this manuscript for U.S.\ Government purposes.
\end{document}